\documentclass[aps,prb,preprint] {revtex4}
\begin{document}
\draft
\newcommand{\lsvo} {Li$_2$VOSiO$_4$ }
\newcommand{\lgvo} {Li$_2$VOGeO$_4$ }
\newcommand{\rat} {J_2/J_1 }
\newcommand{\pmuo} {P_{\mu}(t) }
\newcommand{\lsco} {La$_{2-x}$Sr$_x$CuO$_4$ }
\newcommand{\lscotu} {La$_{1.98}$Sr$_{0.02}$CuO$_4$ }
\newcommand{\lscofri} {La$_{1.97}$Sr$_{0.03}$CuO$_4$ }
\newcommand{\la} {$^{139}$La }
\newcommand{\cu} {$^{63}$Cu }
\newcommand{\cuo} {CuO$_2$ }
\newcommand{\ybco} {YBa$_{2}$Cu$_3$O$_{6.1}$ }
\newcommand{\ybcoca} {Y$_{1-x}$Ca$_x$Ba$_2$Cu$_3$O$_{6.1}$ }
\newcommand{\yt} {$^{89}$Y }
\newcommand{\etal} {{\it et al.} }
\newcommand{\ie} {{\it i.e.} }
\hyphenation{a-long}
%
%
%
%


\title{Very 
low-frequency excitations in frustrated two-dimensional $S=1/2$ Heisenberg antiferromagnets
}
\author{
P. Carretta\footnote{e-mail: carretta@fisicavolta.unipv.it}, R. Melzi and N. Papinutto} 
\address{Dipartimento di Fisica ``A. Volta'' e Unit\'a INFM di Pavia,
Via Bassi 6, 27100 Pavia, Italy} 
\author{
P. Millet}
\address{
Centre d'Elaboration des Mat\'eriaux et d'Etudes Structurales, CNRS, 31055 Toulouse Cedex, 
France
} 
\date{\today}
\widetext
\begin{abstract}





$\mu$SR and $^7$Li NMR relaxation measurements in frustrated two-dimensional $S=1/2$
Heisenberg antiferromagnets on a square lattice are presented. It is found
that both in \lsvo and \lgvo, spin dynamics at frequencies well below
the Heisenberg exchange frequency are present. These dynamics are 
associated with 
the motions of walls separating coexisting collinear domains with a magnetic
wave vector
rotated by 90$^o$.








\end{abstract}
\maketitle
%
\narrowtext



Two-dimensional $S=1/2$ Heisenberg antiferromagnets (2DQHAF) have attracted a lot of interest
in the last decade mainly due to their interplay with high $T_c$ 
superconductors \cite{QHAF}. These
systems are characterized by strong quantum fluctuations in view of the reduced dimensionality
and low spin value, with long range order taking place, in principle, only at zero temperature 
($T$).
Still it is possible to further enhance quantum fluctuations without introducing defects, static
or itinerant, by frustrating the antiferromagnetic exchange interaction.
This situation is encountered in the $J_1$-$J_2$ model on a square lattice, where the next nearest
neighbour interaction $J_2$, along the diagonal of the square, 
competes with the nearest neighbour
interaction $J_1$, along the side of the square \cite{Chandra1,Schulz,Sorella,Chandra2}. 
On increasing $J_2$ one expects first a quantum phase
transition from a N\'eel ordered phase to a non-magnetic ground state, whose precise
nature is still subject of debate \cite{Nuovi},  and then another transition to a collinear ground state.
The collinear phase can be considered as formed by two interpenetrating sublattices with
a reciprocal orientation of the N\'eel vectors which classically can assume any orientation.
However, this degeneracy is lifted by quantum fluctuations and, 
as a consequence of an order from disorder process, just 
two collinear ground-states are realized \cite{Chandra2}: 
collinear I, with magnetic wave-vector ${\bf Q}=(q_x=\pi/a ,q_y=0)$, and collinear 
II, with ${\bf Q}=(0, \pi/a)$.
These two ground
states are degenerate and it is not possible a priori to tell which of one of them
will be realized, unless
other interactions are considered. 


Recently two prototypes of frustrated 2DQHAF with
$\rat\simeq 1$, \lsvo ($\rat= 1.1$ and $J_1+J_2\simeq 8.2$ K) and \lgvo ($\rat= 0.9$ and $J_1+J_2\simeq 6$ K), have been discovered
\cite{Melzi,MelziB}.
\lsvo , which is the most studied
of the two compounds, shows a collinear ground state always of type I,
 with the spins pointing along the $x$ direction. The degeneracy
among the two collinear 
phases is relieved by a structural distortion occuring just above
the transition to the collinear phase at $T_N$
\cite{Melzi,MelziB}. For $J_1+J_2\geq T> T_N$ the system is characterized by 
a correlated spin dynamics with domains of type I and II 
extending over a length of the order of $\xi$, the in-plane correlation
length.



In this manuscript we will provide evidence, based on $\mu$SR and $^7$Li NMR measurements,
that for   $J_1+J_2\geq T> T_N$ very slow spin dynamics are developed in the two systems, with
characteristic frequencies $\omega_c\ll (J_1+J_2)k_B/\hbar$ . 
These dynamics are associated with
the motions of domain walls separating the collinear I and II phases which dynamically coexist 
above $T_N$. New $\mu$SR data showing that both in \lsvo and \lgvo
the transition to the collinear I ground state is possibly triggered by the in-plane XY
anisotropy are also presented.



The measurements were performed on powder samples prepared by solid state
reaction, as described in ref. 9. 
Longitudinal
field (LF) and zero field (ZF) $\mu$SR
measurements were performed at ISIS pulsed source using $29$ MeV/c spin-polarized
muons. The background signal was estimated
below $T_N$, equal to $2.86$ K for \lsvo and $2.1$ K for \lgvo , 
from the amplitude of the slowly decaying 
oscillation in a transverse field of $100$ Gauss. 
For $T<T_N$, the frequency of the oscillating ZF $\mu$SR signal 
directly yields the amplitude of the 
local field
at the muon site $B_{\mu}$, which is proportional to the sublattice magnetization 
(see Fig. 1). 
The $T-$dependence of $B_{\mu}$ in \lgvo is very close to the one
observed in \lsvo \cite{MelziB}. In particular, one finds a critical exponent
for the sublattice magnetization very close to $\beta = 0.235$, the one
predicted for a 2D XY system on a finite size \cite{Bram}. 
The absolute value of $B_{\mu}$,
similar for both compounds, allows to estimate where the $\mu^+$ is
probably localized. By assuming a simple dipolar coupling with V$^{4+}$
magnetic moments, whose absolute value should range between $0.6$ and $0.25$
$\mu_B$ \cite{MelziB}, two possible $\mu^+$ sites have been identified. One around
(Si,Ge)O$_4$ tetrahedral oxygens, the other close to VO$_5$ apical
oxygen.
   

 
Above $T_N$, for
$T\lesssim J_1+J_2$, the decay of the muon 
polarization was observed to be rapidly 
quenched by a LF and to become field-independent above 200 Gauss
(see Fig.2). 
The remanent decay, driven
by high frequency spin fluctuations, is weakly T-dependent and 
should be compared
to NMR spin-lattice relaxation rate $1/T_1$ \cite{MelziB}. 
These observations point out that the dominant contribution
to the decay of $\mu^+$ polarization in ZF is 
not a spin-lattice relaxation process induced
by field fluctuations at frequencies 
$\omega_c\sim (J_1 + J_2)k_B/\hbar$, as we have
erroneously put forward on the basis of ZF measurements only \cite{MelziB}, 
but it is driven by 
very low-frequency fluctuations with $\omega_c\ll (J_1 + J_2)k_B/\hbar$. 
In this case the decay of the polarization is not
described by $\pmuo\propto exp(-\lambda t)$, as for LF
above 200 Gauss, but one has to resort to the complete Kubo-Toyabe function \cite{KT}. 
In the presence of a LF, if the system is not in the purely static or very 
fast fluctuations regime $\omega_c\gg \gamma \sqrt{<\Delta h^2>}$, with
$\gamma= 2\pi\times 13.55$ kHz/Gauss the muon
gyromagnetic ratio and  $\sqrt{<\Delta h^2>}$ the mean square amplitude of field fluctuations,
the fit of $\pmuo$ data with the complete Kubo-Toyabe function is rather 
cumbersome \cite{RDR,Uemura}.
Nevertheless,
if $\omega_c\gtrsim \gamma \sqrt{<\Delta h^2>}$, one 
can resort to the analytical expression
for $\pmuo$ derived by Keren \cite{Keren}, 
which provides a powerful method to fit the decay of muon
polarization in LF. 


For $T\gg J_1 + J_2$ a non-negligible T-independent decay, due to nuclear dipolar coupling,
is still present. Although this contribution might be disregarded for $T\rightarrow T_N$, 
in order to correctly estimate the frequency of these low-energy dynamics also for 
$T\rightarrow J_1+J_2$,
the raw $\mu$SR data of $\pmuo $ have been first divided by 
$\pmuo$ for $T\gg J_1 + J_2$. Then the data have been fitted according to 
\begin{equation}
\pmuo = exp(-\lambda t) \pmuo^{K}(B,\omega_c,<\Delta h^2>) ,
\end{equation}
where
the first term describes the spin-lattice relaxation driven by the fast fluctuations, while the
second one is Keren analytical approximation of Kubo-Toyabe function
(see Eqs. 3 and 4 of Ref. 14), which depends on
the intensity of the longitudinal field $B$, as well as on $\omega_c$ and $<\Delta h^2>$.
By fixing $\lambda$ ($\simeq 0.015 \mu$s$^{-1}$) from 
the high LF measurements above $200$ Gauss and fitting $\mu$SR data at different
magnetic fields, it was possible to estimate $<\Delta h^2>$ and the T-dependence of
$\tau_c\equiv 1/\omega_c$ (see Fig. 3). 
We have found that $\omega_c$ is always at least
$6$ times larger than $\gamma \sqrt{<\Delta h^2>}$, supporting the applicability of Keren
analytical expression. $\tau_c(T)$ is observed to diverge exponentially on decreasing T,
with an effective activation barrier around $3$ K for \lgvo and close to $8$ 
K for \lsvo .
These low frequency fluctuations can originate, in principle, either from $\mu^+$ diffusion
or from a slow spin dynamics. The first possibility seems quite unlikely if one 
considers that in perovskites the activation energies are usually at least two orders of
magnitude larger \cite{Schenck}. In particular, in \lsvo and \lgvo the muon is located either close
to the apical 
oxygen of the VO$_5$ pyramids or close to an oxygen of SiO$_4$ tetrahedra
where it experiences a strong attractive coulomb potential which hinders
its diffusion. On the other hand, the second situation 
is more likely once one realizes that the more rapid increase on cooling
of $\tau_c$ in \lsvo seems related to the larger exchange coupling.


These very slow
fluctuations should cause also an enhancement of $^7$Li NMR transverse
relaxation rate $1/T_2$, which was derived 
from the decay of the echo amplitude after a $\pi/2 -\tau -\pi$ sequence. 
Two main contributions to $1/T_2$ are present. The first is temperature independent and 
is due to nuclear dipole-dipole interaction among $^7$Li nuclei. Its value can be estimated from 
the lattice sums \cite{Slichter} to be $(1/T_2)_{dip}\simeq 5$ ms$^{-1}$. The second one
is a temperature dependent contribution proportional to the
amplitude of the spectral density at $\omega=0$, which sould be very sensitive to these
low-frequency fluctuations \cite{Slichter}. So one can write that
$1/T_2=(1/T_2)_{dip} + (1/T_2)'$, with 
\begin{equation}
(1/T_2)'={\gamma^2\over 2}\int^{\infty}_{-\infty}<h_z(t)h_z(0)> dt =
{\gamma^2\over 2}<\Delta h_z^2>\tau_c .
\end{equation}
Here $h_z(t)$ is the fluctuating component of the local field at $^7$Li nuclei along the external
magnetic field. Now, by assuming $\omega_c\simeq (J_1+J_2)k_B/\hbar$, one would derive a
contribution to $1/T_2$ of at least one order of magnitude smaller than the experimental one.
Moreover, if $\omega_c$ was well above the nuclear Larmor frequency ($\omega_L$), 
$1/T_2$ and $1/T_1$
should show the same T-dependence, while it is evident from Fig. 4 that $1/T_2$ starts diverging
in a T range where $1/T_1$ is still constant.
Only a very low-frequency dynamics at $\omega_c\ll (J_1+J_2)k_B/\hbar$ can explain
the experimental data for $^7$Li NMR transverse relaxation rate. Therefore, the temperature
dependence of $(1/T_2)'\propto \tau_c$ should be characterized by an exponential divergence,
with an effective activation barrier equal to the one estimated from $\mu$SR relaxation
measurements.
In fact, the barrier derived
from $1/T_2$ and $\mu$SR decay rate was found to be the same within $\pm 15$ \% , pointing out that both 
techniques are probing the same dynamics and definetely ruling out the possibility of $\mu^+$ diffusion.
This implies that there are two main dynamics 
contributing to the spectral density: one with characteristic frequencies of the
order of $(J_1+J_2)k_B/\hbar$, probed by NMR $1/T_1$ and $\mu$SR decay rate for $B> 100$ Gauss,
the other with $\omega_c\ll (J_1+J_2)k_B/\hbar$, which is evident in low field $\mu$SR decay rates
and NMR $1/T_2$. 


Since nuclear spin-lattice relaxation rate probes the amplitude of the
spectral density at $\omega_L$, one would expect to observe an enhancement 
in $^7$Li $1/T_1$ on  decreasing the Larmor
frequency towards $\omega_c$. For this
reason $^7$Li NMR $1/T_1$ was measured at $B=3.5$ kGauss 
and compared with the one 
measured at $16$ kGauss. One finds that upon cooling, approaching $T_N$,
an enhancement of $1/T_1$ with decreasing field is evident (see Fig. 4), 
suggesting the presence of very low-frequency dynamics.  These dynamics are not due to spin
diffusion \cite{Boucher}, which in 2D systems gives rise to a logarithmic divergence
of $1/T_1$. First, because if spin diffusion was present one should observe a frequency
dependence also for $T> 3$ K, while it is absent (see Fig. 4b). Second, 
the low-frequency divergence induced by spin-diffusion
should be cut at frequencies corresponding to energies of the order of the spin anisotropy
\cite{Boucher}, which are usually  more than three orders of magnitude larger than $\hbar\omega_c$.


Now, which is the possible origin of the spin dynamics at 
frequencies $\omega_c\ll (J_1+J_2)k_B/\hbar$ ? In frustrated 2DQHAF with $J_1\simeq J_2$ the degeneracy of
the two collinear ground states leads to the coexistence of domains of type I and II
above T$\simeq E(T)$ \cite{Chandra2}, the energy barrier separating the two collinear phases. In 
these vanadates the degeneracy is lifted by a structural distortion occurring just
above $T_N$ \cite{Melzi,MelziB}, so that the double potential well describing the energy levels
becomes asymmetric and the system collapses always in the collinear I phase.
On the other hand, for $J\geq T> T_N$, above the structural distortion,
the double potential well is symmetric and the spectral density is characterized both
by a fast dynamic, at $\omega_c\sim (J_1+J_2)k_B/\hbar $, 
and a slow thermally activated dynamic 
with $\omega_c= \omega_o exp(-E(T) / T)$, $\omega_o$ being a characteristic
attempt frequency.
These low-frequency 
fluctuations can originate from the motions of domain walls separating
the two collinear phases, or, in other terms, from the phase modes
of soliton strings separating domains \cite{Jongh} with a pitch vector ${\bf Q}$
rotated by $90^o$.
Such a low-frequency dynamic is present in one-dimensional QHAF as CuGeO$_3$ 
\cite{Mladen}, where the slowing down of
soliton phase modes manifests itself as an increase of the NMR linewidth. 
$\omega_c$ can be much smaller than $(J_1+J_2)k_B/\hbar$ 
first because of the activated 
type of dynamics, second because $1/\omega_o$ corresponds to the time required
for a domain wall to visit all lattice sites within a domain, which increases 
with the in-plane correlation length $\xi (T)\propto\exp(2\pi\rho_s/T)$, $\rho_s$ being
an effective spin-stiffness \cite{Schultz}.


In order to verify if
the temperature dependence of $\tau_c$ reported in Fig. 3 for \lsvo and \lgvo
is compatible with such a picture, the data were fitted using for $E(T)$ the
expression derived by Chandra et al. \cite{Chandra2}
\begin{equation}
E(T)= \biggl({J_1^2S^2\over 2J_2}\biggr)\biggl[0.26\biggl({1\over S}\biggr) + 
0.318 \biggl({T\over J_2S^2}\biggr)\biggr]\xi^2(T)
\end{equation}
One observes (Fig. 3)
that on approaching $T_N$ the increase of $\tau_c$ can be
well approximated by an activated $T-$dependence, with $E(T)$ given by Eq. 3,
with values for $2\pi\rho_s$ slightly below $(J_1 + J_2)$, as expected for
a frustrated system \cite{Schultz}. In particular,
one finds $2\pi\rho_s\simeq 7.6$ K for \lsvo and $2\pi\rho_s\simeq 4.9$ K for
\lgvo . A departure
from this trend is evident for $T\rightarrow J_1+J_2$ and can be associated 
with the inadequacy of the simple exponential expression used for $\xi(T)$
at $T\simeq J_1+J_2$. The consistency of the $\tau_c(T)$ behaviour with the T-dependence
of $E(T)$ predicted \cite{Chandra2} for the energy barrier separating the two collinear phases,
reinforces our conclusion in favour of a slow dynamics driven by the motions of domain walls.


In conclusion, from the analysis of $\mu$SR and $^7$Li NMR relaxation rates in 
\lsvo and \lgvo it has been shown, for the first time, that in frustrated 2DQHAF,
with $J_1\simeq J_2$, low frequency spin dynamics are present at
frequencies well below the Heisenberg exchange frequency. 
These dynamics are associated with the motions of domain walls
separating the collinear I and II phases which coexist above $T_N$ and are characterized by 
a pitch vector rotated by $90^o$.
 







%
%
%
	
We would like to thank
A. Hillier and J. S. Lord for their help during the  
$\mu$SR measurements at ISIS.











\begin{figure}
\caption{Temperature dependence of the local field at the muon site in \lsvo (circles) 
and \lgvo (squares), normalized by $B_{\mu}(0)= 313$ Gauss and $B_{\mu}(0)= 339$ Gauss,
respectively. The temperature is normalized by $T_N=2.86$ K for \lsvo and by $T_N=2.1$ K for
\lgvo. The dotted line shows the critical behaviour expected for a critical exponent 
$\beta=0.235$.}
\end{figure}


\begin{figure}
\caption{a) Time evolution of the muon polarization in \lsvo in a LF of (from the bottom to
the top) 5, 10, 20, 50, 900 Gauss, for $T=2.95$ K (just above $T_N$). 
In b) $\pmuo$ for a LF of $5$ Gauss is reported
for clarity. The solid lines correspond to the best fits obtained from Eq. 1 with 
$\gamma\sqrt{<\Delta h^2>}=0.65$$\mu$s$^{-1}$ and $\tau_c=0.265$ $\mu$s. 
}
\end{figure}


\begin{figure}
\caption{Temperature dependence of the correlation time for the low-frequency
fluctuations in \lsvo (circles) and in \lgvo(squares). These data have been 
derived from the fit
of $\pmuo$ in a LF of 5 gauss according to Eq. 1, after dividing the raw $\mu$SR data
by $\pmuo$ for $T\gg J_1+J_2$ (see text). The dotted lines show
the best fits according to $\tau_c=(1/\omega_o)exp(E(T)/T)$, with $E(T)$ given by Eq. 3.
The increase in the error bars at high $T$ originates from the increase in the weight
of the dipolar contribution to $\pmuo$ at higher $T$.}
\end{figure}


\begin{figure}
\caption{a) Temperature dependence of $^7$Li NMR $1/T_2$ in \lgvo above $T_N$, for $H= 3.7$ kGauss.
The dotted line shows the best fit according to an activated $T-$ dependence of $1/T_2$
with an effective barrier of $3.5$ K. b) Temperature dependence of $^7$Li NMR $1/T_1$
in \lgvo above $T_N$, for $H=3.7$ kGauss (closed circles) and $H=16$ kGauss (empty circles).}
\end{figure}







\begin{references}




\bibitem{QHAF} see for example E. Dagotto and T. M. Rice, Science 271, 619 (1995)
and E. Dagotto, Rep. Prog. Phys. 62, 1525 (1999) and references therein




\bibitem{Chandra1} P. Chandra and B. Doucot, Phys. Rev. B 38, 9335 (1988)


\bibitem{Schulz} H. J. Schulz and T. A. L. Ziman, Europhys. lett. 18, 355 (1992);


\bibitem{Sorella} S. Sorella, Phys. Rev. Lett. 80, 4558 (1998); L. Capriotti and
S. Sorella, Phys. Rev. Lett. 84, 3173 (2000)


\bibitem{Chandra2} P. Chandra, P. Coleman and A. I. Larkin, Phys. Rev. Lett. 64, 88 (1990)


\bibitem{Nuovi} V. N. Kotov, M. E. Zhitomirsky and O. P. Sushkov, Phys. Rev. B 63, 064412 (2001);
V. N. Kotov, J. Oitmaa, O. P. Sushkov and Z. Weihong; Phys. Rev. B 60, 14613 (1999)


\bibitem{Melzi} R. Melzi et al., Phys. Rev. Lett. 85, 1318 (2000)


\bibitem{MelziB} R. Melzi et al., Phys. Rev. B 64, 024409 (2001)


\bibitem{Millet} P. Millet and C. Satto, Mat. Res. Bull. 33, 1339 (1998)


\bibitem{Bram} S. T. Bramwell and P. C. W. Holdsworth, Phys. Rev. B 49, 8811 (1994);
 S. T. Bramwell and P. C. W. Holdsworth, J. Phys.: Condens. Matter 5, L53 (1993)


\bibitem{KT} R. Kubo, Hyperfine Interactions 8, 731 (1981) 


\bibitem{RDR} see R. De Renzi and S. Fanesi, Physica B 289-290, 209 (2000) and
references therein


\bibitem{Uemura} Y. J. Uemura et al., Phys. Rev. B 31, 546 (1985)


\bibitem{Keren} A. Keren, Phys. Rev. B 50, 10039 (1994)


\bibitem{Schenck} A. Schenck in {\it Muon Spin Rotation: Principles and Applications in 
Solid State Physics} (Hilger, Bristol 1986)


\bibitem{Slichter} C. P. Slichter, in {\it Principles of Nuclear Magnetism}, 
(Springer-Verlag, Berlin 1980)


\bibitem{Boucher} H. Benner and J. P. Boucher, in {\it Magnetic Properties of Layered
Transition Metal Compounds}, edited by L. J. De Jongh (Kluwer Academic, Norwell, MA, 1990), p.323 


\bibitem{Schultz} H. J. Schulz, T. A. L. Ziman and D. Poilblanc, J. Phys. I France 6, 675 (1996)


\bibitem{Jongh} see H. J. M. de Groot and L. J. de Jongh in {\it Magnetic Properties of Layered
Transition Metal Compounds}, Ed. L. J. de Jongh (Dodrecht, Kluwer), p. 379 (1990)


\bibitem{Mladen} G. S. Uhrig et al., Phys. Rev. B 60, 9468 (1999)


\end{references}
\end{document}